\documentclass[twocolumn]{aastex7}

\newcommand{\HI}{H{\,\small I}}

\newcommand{\ltsima} {$\; \buildrel < \over \sim \;$}
\newcommand{\gtsima} {$\; \buildrel > \over \sim \;$}
\newcommand{\lta} {\lower.5ex\hbox{\ltsima}}

\newcommand{\gta} {\lower.5ex\hbox{\gtsima}}

\newcommand{\kms}{km\,s$^{-1}$}

\newcommand{\mjybm}{mJy\,beam$^{-1}$}
\newcommand{\lya}{Ly$\alpha$}

\shorttitle{Molecular Clouds at Cosmic Noon}
\shortauthors{Emonts et al.}
\received{January 6, 2026}
\revised{February 25, 2026}
\accepted{March 5, 2026}
\submitjournal{ApJL}

\begin{document}

\title{Molecular Clouds Resolved at the Onset of Cosmic Noon}

\author[0000-0003-2983-815X]{Bjorn H. C. Emonts}
\affiliation{National Radio Astronomy Observatory, 520 Edgemont Road, Charlottesville, VA 22903, USA}
\email[show]{bemonts@nrao.edu}  

\author[0000-0003-1939-5885]{Matthew D. Lehnert}
\affiliation{Universit\'{e} Lyon 1, ENS de Lyon, CNRS UMR5574, Centre de Recherche Astrophysique de Lyon, F-69230 Saint-Genis-Laval, France}
\email[]{}  

\author[0000-0001-6251-649X]{Mingyu Li}
\affiliation{Department of Astronomy, Tsinghua University, Beijing 100084, China}
\email[]{}  

\author[0009-0007-4080-9807]{Azia Robinson}
\affiliation{Department of Physics and Astronomy, Agnes Scott College, 141 E. College Ave. Decatur, GA, 30030}
\affiliation{National Radio Astronomy Observatory, 520 Edgemont Road, Charlottesville, VA 22903, USA}
\email[]{}  

\author[0000-0002-1163-010X]{Stephen J. Curran}
\affiliation{School of Chemical and Physical Sciences, Victoria University of Wellington, PO Box 600, Wellington 6140, New Zealand}
\email[]{}  

\author[0000-0003-2566-2126]{Montserrat Villar-Mart\'{i}n}
\affiliation{Centro de Astrobiolog\'{i}a, CSIC-INTA, Ctra. de Torrej\'{o}n a Ajalvir, km 4, 28850 Torrej\'{o}n de Ardoz, Madrid, Spain}
\email[]{}  

\author[0000-0001-6647-3861]{Chris L. Carilli}
\affiliation{National Radio Astronomy Observatory, P.O. Box O, Socorro, NM 87801, USA}
\email[]{}  

\author[0000-0002-9482-6844]{Raffaella Morganti}
\affiliation{ASTRON, the Netherlands Institute for Radio Astronomy, Oude Hoogeveensedijk 4, 7991 PD Dwingeloo, The Netherlands.}
\affiliation{Kapteyn Astronomical Institute, University of Groningen, P.O. Box 800, 9700 AV Groningen, The Netherlands}
\email[]{}  

\author[0000-0001-9163-0064]{Ilsang Yoon}
\affiliation{National Radio Astronomy Observatory, 520 Edgemont Road, Charlottesville, VA 22903, USA}
\email[]{}  

\author[0000-0002-2421-1350]{Pierre Guillard}
\affiliation{Sorbonne Universit\'{e}, CNRS UMR 7095, Institut d'Astrophysique de Paris, 98bis bvd Arago, 75014, Paris, France}
\email[]{}  

\author[0000-0003-2884-7214]{George K. Miley}
\affiliation{Leiden Observatory, Leiden University, PO Box 9513, 2300 RA Leiden, The Netherlands}
\email[]{}  

\author[0000-0002-0587-1660]{Reinout J. van Weeren}
\affiliation{Leiden Observatory, Leiden University, PO Box 9513, 2300 RA Leiden, The Netherlands}
\email[]{}  

\author[0000-0001-8467-6478]{Zheng Cai}
\affiliation{Department of Astronomy, Tsinghua University, Beijing 100084, China}
\email[]{}  

%% Use the \collaboration command to identify collaborations. This command
%% takes an optional argument that is either a number or the word "all"
%% which tells the compiler how many of the authors above the command to
%% show. For example "\collaboration[all]{(DELVE Collaboration)}" wil include
%% all the authors above this command.
%%
%% Mark off the abstract in the ``abstract'' environment. 

\begin{abstract}
We present the discovery of seven molecular clouds in the radio galaxy B2\,0902+34 at redshift $z$\,=\,3.4. These clouds are detected as CO(0-1) absorption features against the bright radio continuum, and spectrally resolved using the Karl G. Jansky Very Large Array (VLA). The velocity dispersion of the individual absorption components ranges from 3$-$7 \kms, which is similar to values observed for molecular clouds in the Milky Way and nearby galaxies, and imply cloud radii of $R$\,$\sim$\,10$^{1-2}$\,pc. The absorbing clouds are found in a region of high obscuration inside a 30\,kpc wide stellar nebula, as revealed by rest-frame near-ultraviolet imaging performed with the Hubble Space Telescope (HST). The fact that we spectrally resolve molecular clouds at the onset of Cosmic Noon opens prospects for studying cloud chemistry and physics that drive the formation of stars in the Early Universe.
\end{abstract}

%% Keywords should appear after the \end{abstract} command. 
%% The AAS Journals now uses Unified Astronomy Thesaurus (UAT) concepts:
%% https://astrothesaurus.org
%% You will be asked to selected these concepts during the submission process
%% but this old "keyword" functionality is maintained in case authors want
%% to include these concepts in their preprints.
%%
%% You can use the \uat command to link your UAT concepts back its source.

\keywords{\uat{Radio galaxies}{1343} --- \uat{High redshift galaxies}{734} --- \uat{Molecular gas}{1073} --- \uat{Quasar absorption-line spectroscopy}{1317} --- \uat{Interstellar medium}{847} ---  \uat{Circumgalactic medium}{1879}}

%\keywords{\uat{Damped Lyman-alpha systems}{349} --- \uat{Circumgalactic medium}{1979} --- \uat{Quasar absorption-line spectroscopy}{1317} --- \uat{radio galaxies}{1343} --- \uat{Radio jets}{1347} --- \uat{High redshift galaxies}{734}  --- \uat{protogalaxies}{1298}  --- \uat{Interstellar line absorption}{843}  --- \uat{Interstellar molecules}{849} --- \uat{Star formation}{1569} --- \uat{Protoclusters}{1279}}

%% From the front matter, we move on to the body of the paper.
%% Sections are demarcated by \section and \subsection, respectively.
%% Observe the use of the LaTeX \label
%% command after the \subsection to give a symbolic KEY to the
%% subsection for cross-referencing in a \ref command.
%% You can use LaTeX's \ref and \label commands to keep track of
%% cross-references to sections, equations, tables, and figures.
%% That way, if you change the order of any elements, LaTeX will
%% automatically renumber them.

\section{Introduction}
\label{sec:intro}

Molecular clouds, with temperatures of 10$-$100\,K, are the raw building blocks for the formation of stars. In the Milky Way, molecular clouds can be studied in the (sub-)millimeter regime through line emission from a wide range of molecular species, providing information on cloud chemistry and physics \citep[e.g.,][]{hey15,sah21}. If similar studies can be performed at Cosmic Noon ($z$\,$\sim$\,2$-$3), 
when the density of the cosmic star-formation rate and abundance of the cold neutral gas in the Universe reached their peak \citep{lil13,mad14,cur19}, this will provide insight into the fundamental processes that drive the evolution of galaxies.

However, it is extremely difficult to perform emission-line studies of individual molecular clouds beyond the low-$z$ Universe \citep[see][]{des19}. This is because the typical size of molecular clouds ($\la$10$^{2}$\,pc) corresponds to a few tens of milli-arcsec at $z$\,$\sim$2$-$3, and the flux from emission lines rapidly decreases with distance to the galaxy ($S_{\rm line}$\,$\propto$\,$D_{\rm L}^{-2}$). Therefore, studies of cold molecular gas at high-$z$ typically rely on the integrated emission from large gas reservoirs seen in the brightest lines of simple molecules and atoms, such as carbon monoxide (CO), atomic carbon ([CI]) and ionized carbon ([CII]; e.g.,  \citealt{car13}, \citealt{gen15}, \citealt{des20}). As a result, our knowledge of the building blocks for star formation in the Early Universe remains rudimentary.

A potentially powerful tool for studying molecular clouds in the Early Universe is the detection of molecular gas in absorption against luminous background continuum sources. Because the strength of an absorption line depends on the strength of the background continuum, one can obtain information at much higher spatial resolution and much weaker intrinsic line fluxes compared to using emission lines \citep[see][also \citealt{kro18}, \citealt{bal25}]{com08}. In the millimeter regime, molecular absorption has been detected predominantly at low and intermediate redshifts \citep[e.g.,][see also \citealt{cur11}]{isr90,wik94,ger97,com97,wik18,mac18,all19,com19,ros19,ros24,com24}. Radio galaxy PKS 1830-211 at $z$\,=\,0.89, for which dozens of molecular species were detected, shows the potential of using mm absorption studies to perform astrochemistry \citep{mul11,mul13,mul14,mul21,mul23,mul24}.

The highest redshift millimeter absorber known to date is radio galaxy B2\,0902+34 at $z$\,=\,3.4 \citep{lil88}, detected in CO(0-1) and the dense gas tracer CN(N=0$\rightarrow$1) by \citet[][]{emo24}.\footnote{Only considering molecular absorption against a background radio source. \citet{rie22} detected H$_{2}$O absorption against the Cosmic Microwave Background at $z$\,=\,6.34.} It is also among the highest redshift galaxies in which the \HI\ 21-cm line of neutral gas was detected in absorption  \citep[][see also \citealt{kan07}, \citealt{cur08}, \citealt{adi21}]{uso91,bri93,bru96,cod03,cha04}. The initial detection of CO(0-1) was of low spatial and spectral resolution, revealing little about the nature of the absorption system.

In this Letter, we present a study of the CO(0-1) absorption system in  B2\,0902+34 at high spectral and spatial resolution, using NSF's Karl G. Jansky Very Large Array (VLA) and NASA's Hubble Space Telescope (HST). The spectral resolution of the CO data is sufficient to distinguish individual molecular clouds, while the spatial resolution allows a comparison with the HST imaging to study the clouds' environment. Throughout this Letter, we will assume a cosmology of H$_{0}$\,=\,71 \kms\,Mpc$^{-1}$, $\Omega_{\rm M}$\,=\,0.27, and $\Omega_{\Lambda}$\,=\,0.73 \citep{wri06}. The corresponding angular scale is 7.3\,kpc\,arcsec$^{-1}$.

\section{Data}
\label{sec:data}

The VLA observations were obtained in B-configuration during 20 Jan $-$ 2 Feb 2023 for an on-source exposure time of 16 hours (project 23A-312). The instrumental setup included one subband of 128 MHz with a high spectral resolution of 83.3\,kHz that was centred on the redshifted CO(0-1) absorption line at 26.22\,GHz, as well as seven subbands of 128 MHz with 1 MHz channels that covered the continuum around CO(0-1) from 25.5$-$26.4 GHz. The observations adopted a calibration strategy that was identical to the D-configuration observations described in \citet{emo24}, but with phase calibrator scans every 3\,min. 

The VLA data were calibrated manually using CASA v.6.6.4-34 \citep{casa22}, following the strategy described in \citet{emo24}. A line-data cube was then made from the 83.3\,kHz resolution data by subtracting the continuum using the line-free channels in the (u,v)-domain, imaging the data with robust 0.5 weighting and a channel width of 0.956 \kms, and correcting the signal for the primary beam response. Given that the channel width was much narrower than the line signal, we Hanning smoothed the image cube to an effective velocity resolution of 1.912 \kms\ (i.e., twice the channel width) to increase the signal to noise. A 26\,GHz continuum image was made from the 1\,MHz resolution data by imaging the line-free channels using robust 0.5 weighting, cleaning the signal down to a threshold of 0.5\,\mjybm, and applying a primary beam correction. The line cube and continuum image have a resolution of 0.25$^{\prime\prime}$\,$\times$\,0.22$^{\prime\prime}$ at position angle PA\,=\,80$^{\circ}$. 

Near-infrared imaging was performed with HST on 9 Oct 2023 with an exposure time of 37\,min (project 17268; \citealt{emo24}). We used the F105W wideband filter of the Wide Field Camera 3 (WFC3), which does not include strong emission lines and thus measures the near-ultraviolet (near-UV) continuum emission from starlight across 200\,$-$\,270\,nm in the rest frame. The F105W image was obtained through the Mikulski Archive for Space Telescopes (MAST): \dataset[10.17909/cy88-7e89]{https://doi.org/10.17909/cy88-7e89}.

\section{Results}
\label{sec:results}

\begin{figure*}[htb!]
\centering
\includegraphics[width=0.74\textwidth]{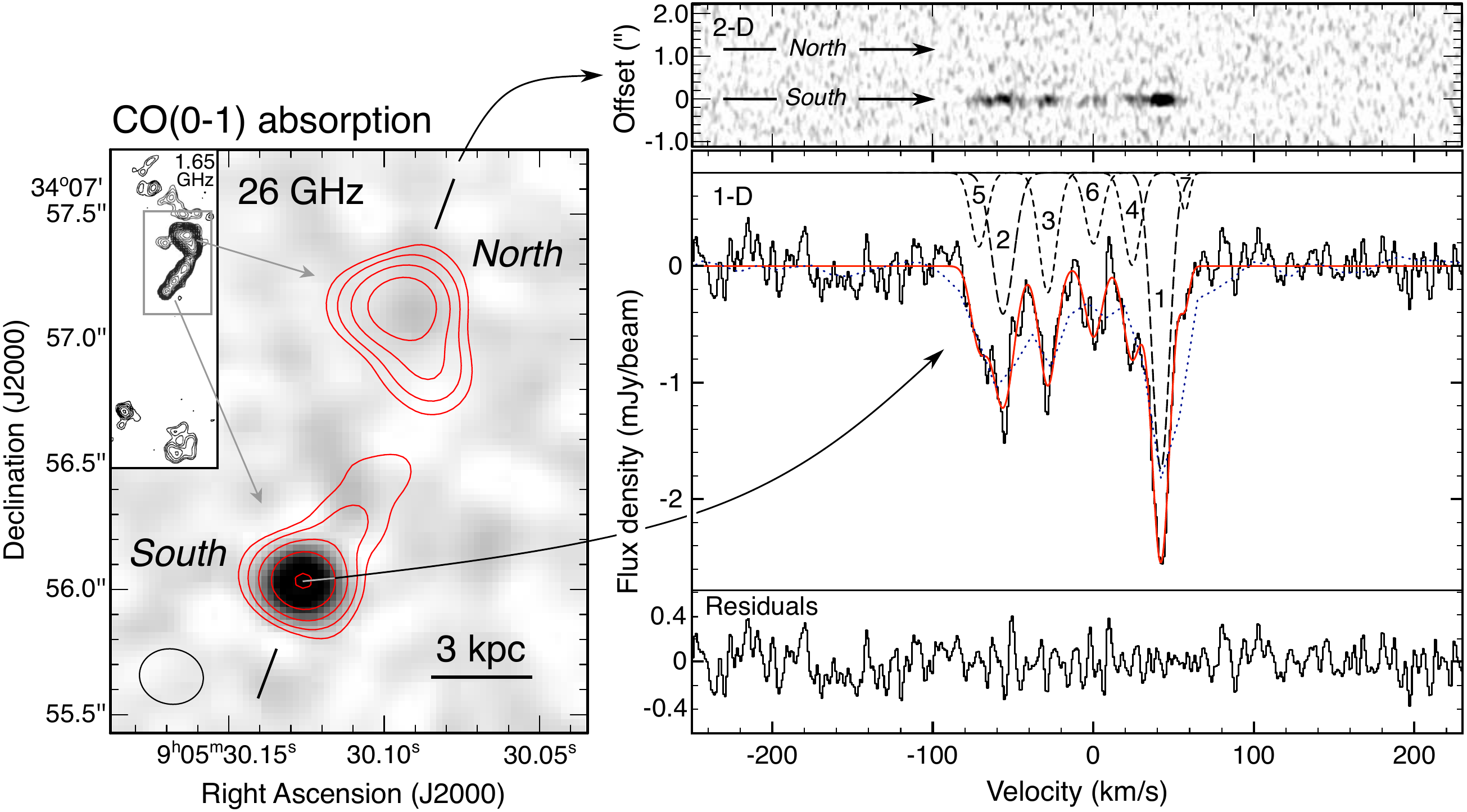}
\caption{{\bf CO(0-1) absorption in B2\,0902+34.} {\sl Left:} Contours of the 26\,GHz radio continuum overlaid onto a total intensity image of the CO(0-1) absorption across the velocity range -80 to 55\,\kms. Contour levels start at 0.5 \mjybm\ and increase by a factor of 2. The inset at the top-left shows the 1.65\,GHz map from \citet{car95}, which reveals additional large-scale emission that our 26\,GHz data do not detect (credit: C.\,Carilli, A$\&$A, 298, 77, 1995, reproduced with permission $\copyright$ ESO). {\sl Right:} CO(0-1) spectrum. The top panel shows a 2-dimensional spectrum along the radio axis, crossing both the southern and northern 26\,GHz component. The molecular clouds seen against the southern component are resolved spectrally, but not spatially. The middle panel shows the 1-dimensional CO(0-1) spectrum taken against the peak of the southern 26\,GHz radio component. The spectrum is Hanning smoothed (Sect. \ref{sec:data}) and has a root-mean-square (rms) noise of 0.15 mJy\,beam$^{-1}$\,chan$^{-1}$ across the line-free channels. The red line represents a combined fit of seven Gaussian components (dashed lines). The blue dotted line is the low-resolution spectrum from \citet{emo24}. The bottom panel shows the residuals after subtracting the model from the spectrum, which are consistent with noise. 
}
\label{fig:COspec}
\end{figure*}

\begin{figure*}[htb!]
\centering
\includegraphics[width=0.78\textwidth]{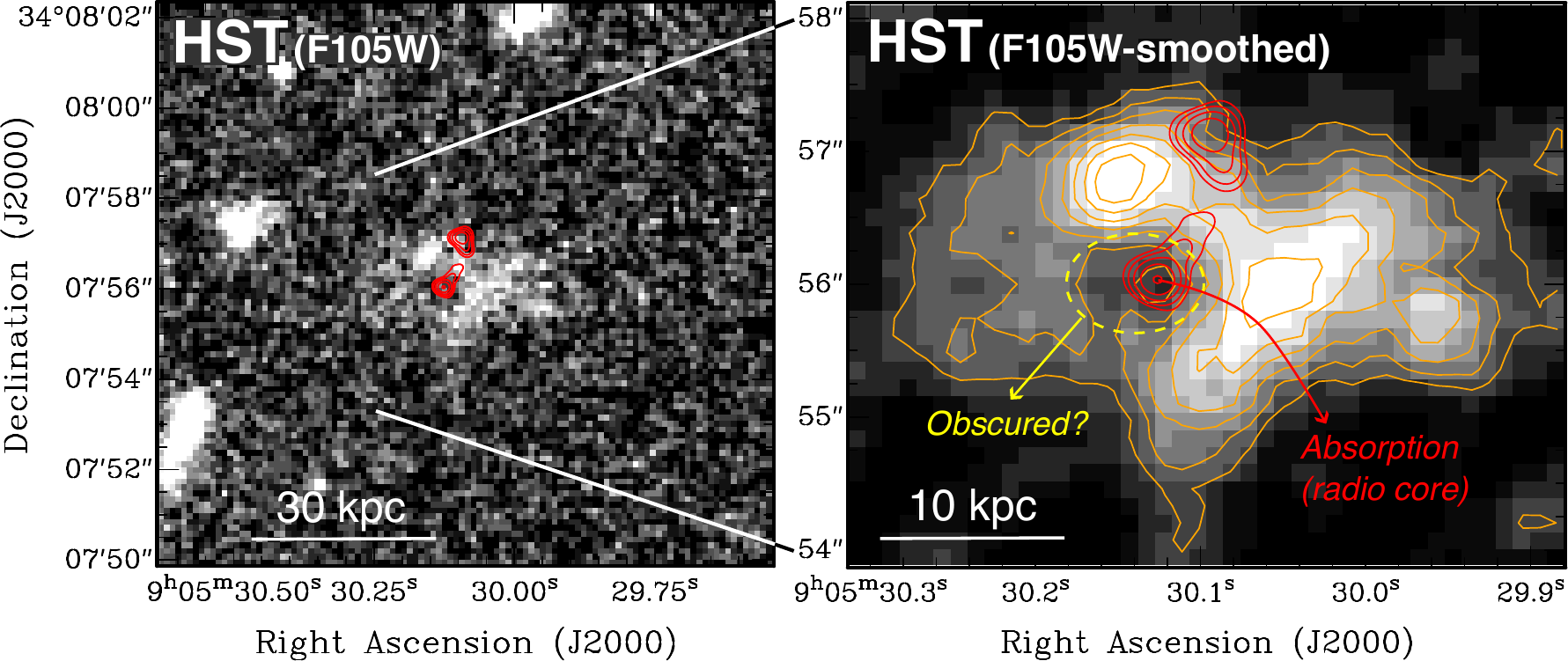}
\caption{{\bf Near-UV starlight around radio galaxy B2\,0902+34.} {\sl Left:} HST/WFC3$_{\rm F105W}$ image of the rest-frame near-UV starlight. The radio source is shown with the same red contours as in Fig.\,\ref{fig:COspec}. {\sl Right:} Zoom-in on the stellar nebula, with a 5-pixel boxcar smoothing applied. The dashed yellow ellipse highlights the region with substantially lower stellar flux, likely due to high obscuration, seen at the location of the southern 26\,GHz radio component. Orange contours show the stellar emission starting at 28$\%$ and increasing by 9$\%$ of the peak intensity of the nebula.
}
\vspace{3mm}
\label{fig:HSTradio}
\end{figure*}

Figure \ref{fig:COspec} (left) shows contours of the 26\,GHz (115\,GHz rest-frame) radio continuum of B2\,0902+34. The radio source shows the same morphology as earlier studies \citep{car94,car95}, consisting of two bright components with fainter emission stretching between them. The northern and southern components have 26\,GHz flux densities peaking at $S_{\rm 26\,GHz}$\,=\,7.9\,$\pm$\,0.4 and 8.4\,$\pm$\,0.4 mJy\,beam$^{-1}$, respectively. We do not detect the faint plume of emission that extends to the north, or the faint emission $\sim$3$^{\prime\prime}$ below the southern component, which were seen at 1.65\,GHz by \citet{car95}.

The CO(0-1) absorption is found against the southern 26\,GHz component and is unresolved at the spatial resolution of our data. The right panel of Fig.\,\ref{fig:COspec} shows the CO(0-1) spectrum. It recovers all the flux from the low-resolution data presented by \citet{emo24}, but the two spectral components that were previously detected are now resolved into seven components, which are fitted with Gaussian functions that have a signal-to-noise ratio between 3 and 33 (see Table \ref{tab:results}). The peak optical depth ($\tau$) of these components ranges from 0.04\,$-$\,0.36. These are likely lower limits, because our beam only partially resolves the complex radio structure seen at lower frequencies \citep{car95}. The velocity dispersion of the components ($\sigma$\,=\,FWHM\,/\,(2$\sqrt{2\,{\rm ln}(2)}$)\,) ranges from 2.7\,$-$\,6.7 \kms.

Figure \ref{fig:HSTradio} shows the radio contours overlaid onto the rest-frame near-UV HST image, which shows diffuse starlight on a scale of $\sim$30 kpc around B2\,0902+34 \citep[see also][]{emo24}. The right panel of Fig.\,\ref{fig:HSTradio} shows that this stellar nebula contains two regions of enhanced emission on either side of the radio source. The morphology and alignment of the stellar nebula matches archival 140\,nm rest-frame HST imaging performed with the F622W filter of the Wide Field and Planetary Camera 2 (WFPC2) \citep[see][]{pen99}, which indicates that the accuracy of the absolute astrometry of these HST images is well within the size of our VLA beam. The southern 26\,GHz radio component, against which we detect the CO absorption, coincides with a region that is particularly devoid of starlight, mimicking a `hole' in the diffuse stellar nebula with a diameter of $\sim$5\,kpc. \citet{car95} concluded that this southern 26\,GHz component hosts the radio core, based on its flat spectral index between 1.65 and 8.3\,GHz. However, we do not detect any galaxy at this location, which would imply that the radio host galaxy is completely obscured, and that the stellar light seen in the HST imaging arises from the circum-galactic environment (see Sect.\,\ref{sec:halo} for more details). Further investigation is needed to verify the location of host galaxy.

\begin{figure*}[htb]
\centering
\includegraphics[width=\textwidth]{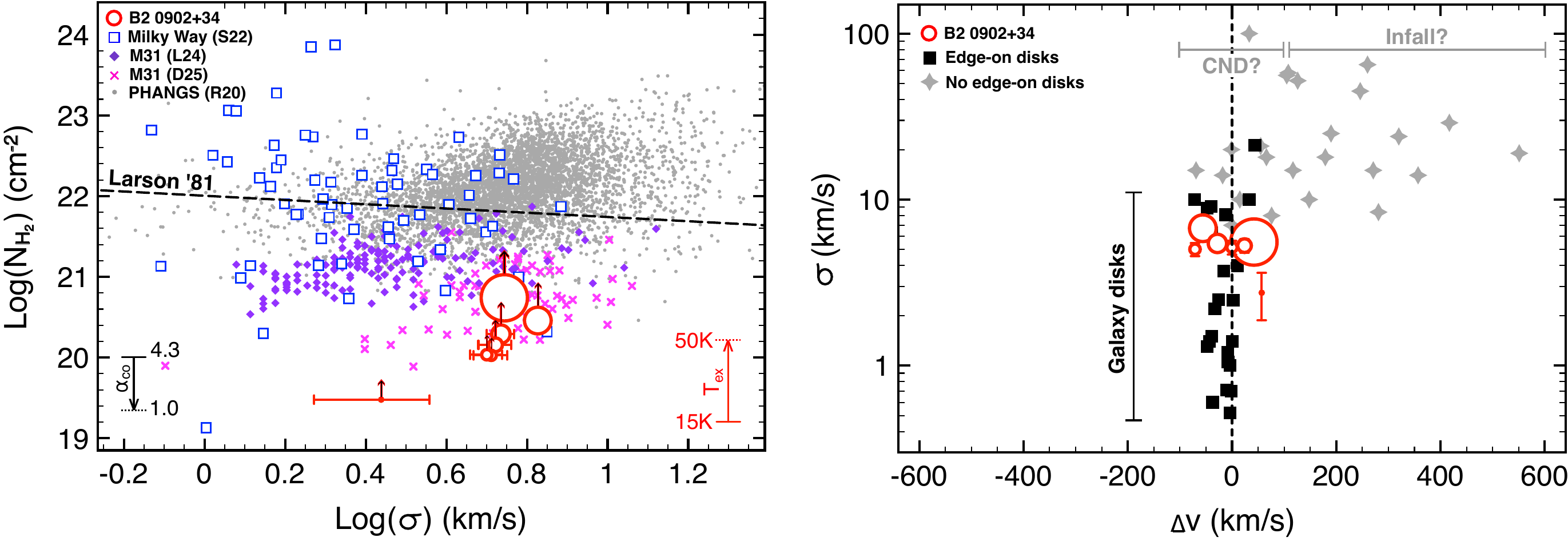}
\caption{{\bf Comparison to molecular clouds in the nearby Universe.} {\sl Left:} H$_{2}$ column density plotted against velocity dispersion for molecular clouds detected in low-$J$ CO($J$+1,\,$J$) emission in the Milky Way Galaxy \citep[][S22]{spi22}, M31 (\citealt{lad24}, L24; \citealt{den25}, D25), and nearby galaxies of the PHANGS (Physics at High Angular resolution in Nearby GalaxieS) survey \citep[][R21]{ros21}. For the emission-line work, $N_{\rm H_2}$ is calculated from the cloud mass ($M$) and radius ($R$) as an average over a projected circular area, assuming a uniform volume density and spherical cloud geometry, following $N_{\rm H_2}$\,=\,$M$/($\pi R^{2}$). All mass values are scaled to a mass-to-light ratio of $\alpha_{\rm CO}$\,=\,4.3 \citep{bol13}, with the PHANGS result including an additional dependence on local metallicity \citep{ros21}. The red circles show the CO(0-1) absorption components of B2\,0902+34, with their size scaled to the integrated line flux (Table\,\ref{tab:results}). The dashed black line shows the correlation derived from Larson's scaling relations between $\sigma$, $R$, and $M$ \citep{lar81}. The arrow at the bottom left [right] corner shows how the emission [absorption] data decrease [increase] in value when $\alpha_{\rm CO}$ [$T_{\rm ex}$] changes from 4.3 to 1.0 [15\,K to 50\,K]. {\sl Right:} Velocity dispersion plotted against the velocity offset from the systemic redshift for molecular absorption components in nearby galaxies, for which also the molecular emission is observed (from \citealt{ros24}, including data from \citealt{ros20} and \citealt{wik97}). Black squares are absorption components found in galaxies with large-scale, edge-on (45$^{\circ}$\,$<$\,$i$\,$\le$\,90$^{\circ}$) molecular disks. Gray stars are absorption features in galaxies lacking an edge-on disk, and likely represent a combination of circum-nuclear disks (CND) and infalling gas \citep[see][]{ros24}. Red circles show the absorption in B2\,0902+34.}
\label{fig:sample_comparison}
\end{figure*}

\begin{deluxetable*}{ccccccccc}
\label{tab:results}
\tablewidth{0pt}
\tablecaption{Measurement of the observed CO(0-1) absorption components.}
\tablehead{
\colhead{Component} & 
\colhead{$z$} & 
\colhead{$\Delta$v$^{*}$} & 
\colhead{$\sigma$} & 
\colhead{$S_{\rm peak}$} & 
\colhead{$\tau_{\rm obs}$} &
\colhead{$\int_{\rm v}\tau_{\rm obs}\,\delta$v}  &
\colhead{$N_{\rm H_2}$\,$^{\dagger}$} &
\colhead{S/N$^{\ddagger}$}
\ \\
\colhead{} &
\colhead{} &
\colhead{(\kms)} &
\colhead{(\kms)} &
\colhead{(mJy)} &
\colhead{} &
\colhead{(\kms)} & 
\colhead{($\times$10$^{20}$ cm$^{-2}$)} &
\colhead{} 
}
\startdata
1 & 3.3966\,$\pm$\,0.0001 & 42\,$\pm$\,1  & 5.5\,$\pm$\,0.2 & -2.54\,$\pm$\,0.06 & 0.36\,$\pm$\,0.02 & 5.0\,$\pm$\,0.3  & 5.5\,$\pm$\,0.3 & 33 \\
2 & 3.3952\,$\pm$\,0.0001 & -56\,$\pm$\,1 & 6.7\,$\pm$\,0.4 & -1.21\,$\pm$\,0.06 & 0.16\,$\pm$\,0.01 & 2.6\,$\pm$\,0.2 & 2.9\,$\pm$\,0.2 & 18 \\
3 & 3.3956\,$\pm$\,0.0001 & -28\,$\pm$\,1  & 5.4\,$\pm$\,0.4 & -1.03\,$\pm$\,0.06 & 0.13\,$\pm$\,0.01 & 1.8\,$\pm$\,0.2 & 1.9\,$\pm$\,0.2 & 13 \\
4 & 3.3963\,$\pm$\,0.0001 & 24\,$\pm$\,1 & 5.3\,$\pm$\,0.5 & -0.79\,$\pm$\,0.06 & 0.10\,$\pm$\,0.01 & 1.3\,$\pm$\,0.2 & 1.4\,$\pm$\,0.2 & 10 \\
5 & 3.3949\,$\pm$\,0.0001 & -71\,$\pm$\,1  & 5.0\,$\pm$\,0.5 & -0.63\,$\pm$\,0.06 & 0.08\,$\pm$\,0.01 & 1.0\,$\pm$\,0.1 & 1.1\,$\pm$\,0.1 & 7.9 \\ 
6 & 3.3960\,$\pm$\,0.0001 & 0\,$\pm$\,1 & 5.1\,$\pm$\,0.5 & -0.61\,$\pm$\,0.05 & 0.08\,$\pm$\,0.01 & 1.0\,$\pm$\,0.1 & 1.1\,$\pm$\,0.1 & 7.7 \\
7 & 3.3968\,$\pm$\,0.0001 & 57\,$\pm$\,1  & 2.7\,$\pm$\,0.9 & -0.33\,$\pm$\,0.08 & 0.04\,$\pm$\,0.01 & 0.3\,$\pm$\,0.1 & 0.3\,$\pm$\,0.1 & 3.0 \\
\enddata
\tablecomments{
$^{*}$ Velocity relative to the systemic redshift $z$\,=\,3.3960 inferred by \citet{emo24}. Because the exact systemic redshift is uncertain, we also list the redshifts of all components in the Table, even though the differences in $z$ are due to relative motions of the clouds rather than cosmological effects.\\
$^{\dagger}$ Following \citet[][their Sect. 4.1.1]{emo24}, assuming $T_{\rm ex}$\,=\,15\,K \citep{wil97} and $N_{\rm CO}$/$N_{\rm H_2}$\,=\,1\,$\times$\,10$^{-4}$ \citep{fre82}. See Appendix \ref{app:column} for details.  \\
$^{\ddagger}$ Signal-to-noise ratio (significance) of the integrated absorbing flux, following \citet[][their Equation\,2]{emo14}. }
\end{deluxetable*}

\section{Discussion}
\label{sec:discussion}

We used the VLA to spectrally resolve the complex absorption system in high-$z$ radio galaxy B2\,0902+34, and used HST to study its environment. In this Section, we will investigate the properties and environment of the molecular gas traced by the absorption.

\subsection{Molecular Clouds}
\label{sec:GMCs}

In Figure \ref{fig:sample_comparison} (left) we compare the observed properties of the CO(0-1) absorption components from Fig.\,\ref{fig:COspec} with the properties of individual molecular clouds observed in CO(1-0) or CO(2-1) emission in the Milky Way and nearby galaxies. For B2\,0902+34 we can directly measure the velocity dispersion, which ranges from $\sigma$\,$\sim$\,3$-$7 \kms\ for the various absorption components. This is consistent with CO velocity dispersions measured for individual molecular clouds in the nearby Universe. The observed optical depths (Table\,\ref{tab:results}) can be used to estimate the average column density for each absorption component following \citet[][their Equation\,1]{emo24}; see Appendix~\ref{app:column}. As summarized in Table\,\ref{tab:results}, this results in $N_{\rm H_2}$\,$\sim$\,(0.3$-$5.5)\,$\times$\,10$^{20}$ cm$^{-2}$, assuming a conservative excitation temperature of $T_{\rm ex}$\,=\,15\,K \citep{wil97}. These values of $N_{\rm H_2}$ are strictly lower limits, because the integrated optical depth is a lower limit (Sect.\,\ref{sec:results}) and because a higher temperature increases the expected column density. For example, when the cloud temperature is $T_{\rm ex}$\,=\,50\,K, the estimated column density increases by a factor of ten. Despite differences between the samples of galaxies in Fig.\,\ref{fig:sample_comparison}, the similarity of $\sigma$ and $N_{\rm H2}$ when comparing the absorption in B2\,0902+34 with values observed in molecular clouds the Milky Way and nearby galaxies suggests that the CO components resolved by our VLA data likely arise from individual molecular clouds at redshift $z$\,=\,3.4. Although this interpretation is somewhat simplistic, our results show the prospect of studying variations in physical or chemical properties among molecular clouds in B2\,0902+34.

Based on $\sigma$, we can get a rough estimates of the cloud diameter $L$ and cloud mass $M$ by following Larson's scaling relations: $\sigma$\,(\kms)\,=\,1.10\,$L$\,(pc)$^{\,0.38}$ and $\sigma$\,(\kms)\,=\,0.42\,$M$\,(M$_{\odot})^{\,0.2}$ \citep{lar81}. These relations were confirmed in various forms by other observational studies \citep[e.g.,][]{sol87,hey09,lom10,miv17,spi22,cho24} and supported by theory \citep{fie11,mck89,kri13}. As shown in Appendix~\ref{app:radius}, following Larson's relations, we expect the molecular clouds in B2\,0902+34 (Table\,\ref{tab:results}) to have a typical radius of roughly $R$\,$\sim$\,10$^{1-2}$ pc  and molecular gas mass of $M$\,$\sim$\,10$^{5-6}$\,M$_{\odot}$.

\subsection{Environment of the Clouds}
\label{sec:environment}

To better understand the nature of the absorbing clouds, we compare our results to \citet{ros24}, who analyzed molecular absorption lines together with the global molecular emission-line properties in eight low-$z$ galaxies with a radio source. As shown in Fig.\,\ref{fig:sample_comparison} (right), they identified distinct populations of absorbing clouds; those with a low velocity dispersion ($\sigma$\,$\la$\,10 \kms) and low relative velocity compared to the systemic velocity ($|\Delta$v$|$\,$\la$\,100\,\kms) are linked to galaxies with large edge-on disks of cold molecular gas, while those with a high velocity dispersion ($\sigma$\,$\sim$\,10\,$-$\,100 \kms) and a high relative velocity ($\Delta$v\,$\la$\,600 \kms) are likely associated with infalling gas. For B2\,0902+34, values for $\Delta$v and $\sigma$ (Table\,\ref{tab:results}) are well within the range of the values that \citet{ros24} associate with disk rotation, not with infall. Therefore, even though B2\,0902+34 is less evolved than these low-$z$ galaxies and contains a more powerful radio source, the observed CO kinematics suggest that the clouds are likely part of a relatively settled reservoir of gas.

To distinguish whether the clouds occupy the region near the Active Galactic Nucleus (AGN) or occur on larger scales, we consider that the absorption occurs in a region that appears to be heavily obscured across $\sim$5\,kpc in the HST imaging (Fig.\,\ref{fig:HSTradio}). It is likely that the absorbing clouds are part of the gas and dust reservoir that causes this large-scale obscuration, rather than that they occupy only a small region surrounding the AGN. As we will discuss in the next section, we speculate that this gas reservoir could be the interstellar medium of the radio host galaxy, which is obscured and hidden within the stellar nebula. Alternatively, the absorbing clouds could be part of a dusty region within the circum-galactic medium (CGM).

\subsection{HST-dark Radio Galaxy in Stellar Halo?}
\label{sec:halo}

The stellar emission that we detect at low surface brightness around B2\,0902+34 is distributed across the central 30\,kpc region of the giant ($\sim$80\,kpc) \lya\ halo of warm neutral/ionized gas seen by \citet{reu03}. It is unlikely that this stellar nebula represents a single large galaxy, given that infra-red observations of other high-z radio galaxies revealed clumpy structures with relatively small ($<<$30\,kpc) stellar bodies hosting the radio source \citep[e.g.,][]{pen01,per24,roy24,sax25,zha26}. Instead, the stellar nebula appears morphologically similar to the halo of diffuse starlight and merging proto-cluster galaxies seen around the Spiderweb Galaxy \citep{mil06}, albeit with less structural details due to a 20$\times$ shorter exposure time. In the Spiderweb, the region spanned by the stellar nebula displays in-situ star formation across the circum-galactic medium \citep{mil06,hat08,emo16}, as predicted by simulations \citep{ahv24}, and will likely evolve into a single massive cluster galaxy \citep{pen07,mil06,hat08,hat09}. While further analysis of the stellar nebula of B2\,0902+34 is beyond the scope of this Letter, the similarity between the stellar nebulae in B2\,0902+34 and the Spiderweb supports the idea proposed in earlier work that B2 0902+34 is likely a young system on its way to forming a massive galaxy \citep{eis92,eal93,pen99}.

The CO absorption is observed in a region of the nebula that is devoid of stellar emission (Fig.\,\ref{fig:HSTradio}). Given that this is the likely location of the radio core (Sect.\,\ref{sec:results}), we argue that the radio host galaxy itself could be heavily obscured, or ``HST dark'', despite the bright stellar emission in its circumgalactic environment. HST-dark galaxies are massive galaxies at $z$\,$>$\,3 that are heavily obscured by dust and which are too faint to be detected with HST, with a visual extinction of $A_{V}$\,$\sim$\,2 mag \citep{bar23}. If the radio host galaxy of B2\,0902+34 fits this definition, it would add credibility to the hyphothesis by \citet{cur21} that high-redshift millimeter absorbers prodominantly occur in systems with a high degree of reddening due to dust, which have largely evaded detection due to selection biases at optical wavelengths \citep[see also][]{cur11,emo24}.

We can estimate the expected obscuration along the sight-line towards the radio source {in B2\,0902+34} based on the total hydrogen column density, assuming a typical gas-to-dust ratio \citep[e.g.,][]{boh78,rac09,shu21}. This total hydrogen column density includes molecular hydrogen (H$_{2}$) and neutral hydrogen (\HI), both of which are seen in absorption in B2\,0902+34 (Sect.\,\ref{sec:intro}; see also figure 4 in \citealt{emo24}). Our sight-line towards B2\,0902+34 captures an integrated H$_{2}$ column density of $N_{\rm H_2}$\,$\ga$\,1.4\,$\times$\,10$^{21}$\,cm$^{-2}$ (Table\,\ref{tab:results}). Estimates for the \HI\ column density derived from the observed \HI\ absorption profile range from $N_{\rm HI}$\,$\sim$\,1.6\,$\times$\,10$^{21}$\,cm$^{-2}$ \citep{cha04} to $N_{\rm HI}$\,$\sim$\,3\,$\times$\,10$^{21}$ \citep{cod03}, based on an assumed spin temperature of $T_{\rm spin}$\,=\,1000\,K. Taking the conservative estimate of $N_{\rm HI}$\,$\sim$\,1.6\,$\times$\,10$^{21}$\,cm$^{-2}$ from \citet{cha04}, the lower limit for the total amount of hydrogen atoms along our line-of-sight is then $N_{\rm H}$\,$\equiv$\,$N_{\rm HI}$\,+\,2$N_{\rm H_2}$\,$\ga$\,4.4\,$\times$\,10$^{21}$\,cm$^{-2}$. Following \citet{shu21}, this corresponds to an average reddening of $E(B-V)$\,$\ga$\,0.7, which is larger than values observed along most sight-lines through the disk of the Milky Way, and similar to the large reddening observed in infrared-luminous starburst galaxies \citep{pog00}. A reddening of $E(B-V)$\,$\ga$\,0.7 can be translated into $A_{V}$\,=\,$R_{V}$\,$\times$\,$E(B-V)$, with the ratio $R_{V}$ ranging from 2.6 to 5 \citep[e.g.,][]{cla88,cal00}, with a typical value of $R_{V}$\,$\sim$\,3.1 across a range of environments \citep[e.g.,][]{rie85,sch98}. Adopting $R_{V}$\,$\sim$\,3.1 translates into $A_{V}$\,$\ga$\,2.2 mag for the absorption region in B2\,0902+34, consistent with values observed in HST-dark galaxies \citep{bar23}. 

B2\,0902+34 is one of very few distant radio galaxies detected in CO mm-absorption, and also the only known high-$z$ radio galaxy where the main stellar body of the radio host appears to be fully obscured in HST imaging. This makes it an interesting target for future infrared and submillimeter observations, to verify the location of the host galaxy and determine whether B2\,0902+34 is the first HST-dark {\sl radio} galaxy.

\section{Concluding remarks}
\label{sec:concluding}

We have used the VLA to spectrally resolve molecular clouds in radio galaxy B2\,0902+34 at $z$\,=\,3.4. The strong background continuum and high obscuration of this system offer a unique opportunity to detect molecular lines in absorption. Future observations, such as with the Next-Generation VLA \citep{mur18}, will be able to spatially resolve the background continuum at a resolution that matches the sizes of individual molecular clouds (i.e., tens of milli-arcsec). This will allow measurements of the distribution and intrinsic optical depth of the absorbing clouds for many different molecular species. B2\,0902+34 therefore offers the exciting prospect of investigating the chemical composition and physical properties of cold molecular clouds at the onset of Cosmic Noon.

\begin{acknowledgments}
We thank the anonymous referee for insightful feedback that improved this paper. As part of this work, AR was a student at the National Radio Astronomy Observatory. The National Radio Astronomy Observatory and Green Bank Observatory are facilities of the U.S. National Science Foundation operated under cooperative agreement by Associated Universities, Inc. Based on observations with the NASA/ESA Hubble Space Telescope obtained from the Data Archive at the Space Telescope Science Institute, which is operated by the Association of Universities for Research in Astronomy, Incorporated, under NASA contract NAS5-26555. Support for program numbers HST-GO-16891 and HST-GO-17268 was provided through a grant from the STScI under National Aeronautics and Space Administration (NASA) contract NAS5-26555. MVM acknowledges support from grant PID2021-124665NB-I00 by the Spanish Ministry of Science and Innovation (MCIN) / State Agency of Research (AEI) / 10.13039/501100011033 and by the European Regional Development Fund (ERDF) “A way of making Europe”. RJvW acknowledges support from the ERC Starting Grant ClusterWeb 804208.
\end{acknowledgments}

\facilities{VLA, {\it HST}}
\software{{CASA, IRAF, KARMA}}

\appendix

\section{H$_{2}$ Column Densities}
\label{app:column}

Estimates of the H$_{2}$ column densities ($N_{\rm H_2}$) shown in Table\,\ref{tab:results} have been derived from the CO column densities ($N_{\rm CO}$), assuming a conversion of $N_{\rm CO}$/$N_{\rm H_2}$\,=\,1\,$\times$\,10$^{-4}$ \citep{fre82}. The derivation of $N_{\rm CO}$ has been extensively summarized in the literature \citep[see, e.g.,][]{wik95,wil13,man15,all19,ros19}. As in \citet{emo24}, we here follow \citet{all19}: 
\begin{equation}
N_{\rm CO} \ge \frac{8\pi}{c^3} \frac{{\nu}^{3}}{g_{J+1} A_{J+1}} \frac{Q(T_{\rm ex})e^{E_{J}/k_{B}T_{\rm ex}}}{1-e^{-h\nu/k_{B}T_{\rm ex}}} \int \tau_{\rm CO} \delta v 
\end{equation}
For the CO(0-1) transition, $\nu$\,=\,115.27\,GHz is the rest frequency of the line, $g_{J+1}$ is the statistical weight for the upper ($J+1$\,=\,1) energy level, $Q({T_{\rm ex})}$ is the partition function based on a single excitation temperature $T_{\rm ex}$, $E_{J}$ is the energy of the lower level $J$\,=\,0, $h$ is the  Planck constant, and $k_{B}$ is the Boltzmann constant. Following \citet{all19}, we can use the rigid-rotator approximation for CO to adopt $g_{J+1}$\,=\,2$(J+1)$+1, $Q$($T_{\rm ex}$)\,$\approx$\,$T_{\rm ex}$/B (for $T_{\rm ex}>>B$), and $E_{J}$/$k_{b}$\,$\approx$\,$J$($J+1$)$B$. $B$\,=\,2.766\,K is the rotational constant for CO, while the Einstein A-coefficient for CO(0-1) is $A_{J+1}$\,=\,7.67$\times$10$^{-8}$ s$^{-1}$ \citep{cha96}. $\int\tau_{\rm CO}\delta v$ is the integrated optical depth from Table \ref{tab:results}. Assuming $T_{\rm ex}$\,=\,15\,K, the corresponding H$_{2}$ column densities of the individual absorption components are summarized in Table\,\ref{tab:results}.

\section{Radius and Mass Estimates}
\label{app:radius}

To estimate the radius ($R$) and mass ($M$) of the molecular clouds based on their velocity dispersion ($\sigma$), we follow Larson's scaling relations, derived from CO observations of Milky Way clouds \citep{lar81}:
\begin{equation}
\sigma\,(\rm km\,s^{-1})\,=\,1.10\,\cdot\,L\,(pc)^{\,0.38}
\label{eq:L}
\end{equation}
\begin{equation}
\sigma\,(\rm km\,s^{-1})\,=\,0.42\,\cdot\,M\,(M_{\odot})^{\,0.2} 
\label{eq:M}
\end{equation}
For these relations, $L$\,=\,2$R$ is the total linear extent (diameter) and $M$ is the total mass of the molecular cloud, assuming that 70$\%$ is in the form of H$_{2}$. Estimated values based on these scaling relations are given in Table \ref{tab:estimates}. The rms deviation of log($\sigma$) is 0.14 for the size relation and 0.12 for the mass relation \citep{lar81}. The corresponding rms error in $\sigma$ is added in quadrature to the fractional measurement error in $\sigma$ quoted in Table\,\ref{tab:results}, and translated into a an error in log($\sigma$) for each cloud. After re-writing the Equations \ref{eq:L} and \ref{eq:M} to have $R$ and $M$ as dependent variables ($R$\,=\,0.39\,$\cdot$\,$\sigma^{2.63}$ and $M$\,=\,76.5\,$\cdot$\,$\sigma^{5}$), we then propagate this error into an estimated error of log($M$) and log($R$) for each cloud. The corresponding uncertainties in $R$ and $M$ are provided in Table \ref{tab:estimates}. We stress that $R$ and $M$ are not direct measurements for the molecular clouds detected in absorption, so should only be used as an indication of the likely cloud properties in B2\,0902+34.

\begin{deluxetable}{ccc}
%\digitalasset
\label{tab:estimates}
\tablewidth{0pt}
\tablecaption{Estimated radii and masses of the molecular clouds in B2\,0902+34, based on the scaling relations by \citet{lar81}.}
\tablehead{
\colhead{Component} &
\colhead{$R$} &
\colhead{$M$} 
\ \\
\colhead{} &
\colhead{(pc)} &
\colhead{($\times$10$^{5}$\,$M_{\odot}$)} 
}
\startdata
1 & 35$^{+48}_{-20}$   & 4.0$^{+12}_{-3.0}$ \\
2 & 58$^{+80}_{-34}$   & 10$^{+32}_{-7.9}$ \\
3 & 34$^{+47}_{-20}$   & 3.6$^{+12}_{-2.8}$  \\
4 & 31$^{+44}_{-18}$   & 3.1$^{+10}_{-2.4}$  \\
5 & 27$^{+38}_{-16}$   & 2.4$^{+8.0}_{-1.9}$ \\
6 & 29$^{+41}_{-17}$   & 2.8$^{+9.2}_{-2.1}$  \\
7 & 5.5$^{+12}_{-3.9}$     & 0.12$^{+0.86}_{-0.10}$  \\
\enddata
%\tablecomments{}
\end{deluxetable}

%\bibliography{sample7}{}
%\bibliographystyle{aasjournalv7}

%% This command is needed to show the entire author+affiliation list when
%% the collaboration and author truncation commands are used.  It has to
%% go at the end of the manuscript.
%\allauthors

%% Include this line if you are using the \added, \replaced, \deleted
%% commands to see a summary list of all changes at the end of the article.
%\listofchanges

\end{document}